# Achieving very long lifetimes in optical lattices with pulsed cooling

Michael J. Gibbons, Soo Y. Kim, Kevin M. Fortier, Peyman Ahmadi, and Michael S. Chapman

*School of Physics, Georgia Institute of Technology, Atlanta, Georgia 30332-0430*

We have realized a one dimensional optical lattice for individual atoms with a lifetime >300 s, which is 5× longer than previously reported. In order to achieve this long lifetime, it is necessary to laser cool the atoms briefly every 20 s to overcome heating due to technical fluctuations in the trapping potential. Without cooling, we observe negligible atom loss within the first 20 s followed by an exponential decay with a 62 s time constant. We obtain quantitative agreement with the measured fluctuations of the trapping potential and the corresponding theoretical heating rates.

PACS numbers: 37.10.Jk, 37.10.De, 37.10.Vz

Optical dipole traps have become an essential tool in ultracold atomic and molecular physics since the first demonstration in 1986 [1]. They have applications in important research areas including atomic frequency standards [2], tests of fundamental symmetries [3], quantum degenerate gases [4, 5], and development of scalable quantum information processing systems [6]. Optical trapping potentials can be tailored through the choice of optical wavelengths and laser beam configurations to yield a wide variety of trapping arrangements. Optical lattices are particularly useful for confining neutral atom qubits with sub-wavelength precision [7] and for studying quantum phase transitions in quantum gases [8].

Early work with optical dipole traps focused on overcoming short trap lifetimes and excessive atom heating in the traps. These limitations are due to both fundamental heating mechanisms associated with trap light absorption by the atoms as well as heating mechanisms due to fluctuating trapping forces. The far off-resonant trap (FORT) was developed to essentially eliminate fundamental lifetime and heating limits by utilizing large trap light detunings [9]. Nonetheless, early efforts to create ultracold atoms in FORTs were plagued by higher than anticipated heating rates. It was pointed out that fluctuations in the trap potential due to laser intensity noise and/or pointing instabilities can cause heating that limits the lifetime and temperature of optically trapped atoms [10, 11]. By using sufficiently quiet lasers, very long lifetime traps were observed [12, 13], and cooling to quantum degeneracy directly in an optical trap was achieved [4].

With the advent of very stable high power diode and fiber lasers, quantum degenerate gases are now routinely created and studied in optical traps. Surprisingly, beyond the general appreciation that technical heating can limit the performance of optical traps, there has been little attempt to quantitatively study the lifetime limits due to heating in experiments. This is in part because observing any lifetime limits due to technical heating in a FORT made with today's low noise lasers requires exceptional vacuum conditions (<$10^{-10}$ Torr) in order to overcome lifetime limits due to background collisions. One exception is the study by Alt and coworkers [14], who showed that their 3 s trap lifetime was limited by noise applied to their trap beams via an acousto-optic modulator (AOM) used to control the laser beam.

It is particularly important to study heating limitations in optical lattices, since the heating rates scale strongly with the trap frequencies [10], which are typically much higher in optical lattices. Indeed, while trap lifetimes exceeding 300 s have been observed in low frequency optical traps [13], the longest reported lifetime in an optical lattice is no more than 60 s [15].

In this report, we examine heating sources and quantify their effects on the trap lifetime in an optical lattice. We demonstrate a lattice lifetime >300 s by briefly laser cooling the trapped atoms every 20 s to counteract heating. Without the cooling, we measure a non-exponential decay of the atoms with an asymptotic trap lifetime of 62 s. We accurately measure the trap heating sources and obtain good quantitative agreement of the heating rates to numerical simulations.

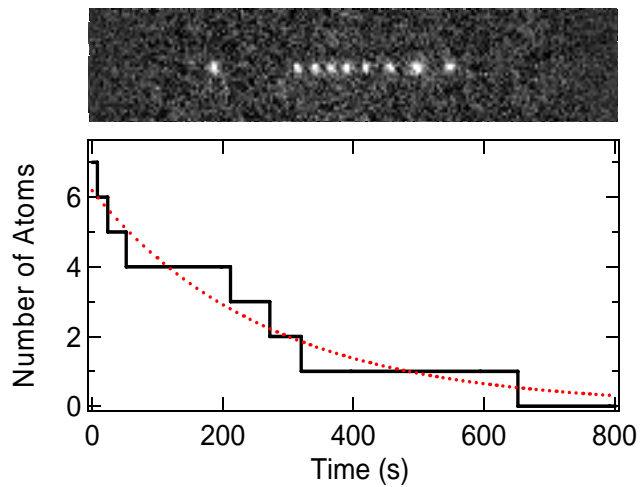

FIG. 1: (top) Continuously cooled array of $^{87}$Rb atoms trapped in a 1-D optical lattice. (bottom) Number of atoms vs. time, starting from seven atoms initially loaded in the lattice. Dashed line is an exponential fit to the data.

The experiment is performed in a vacuum system consisting of a 27×27×150 mm rectangular glass cell attached to a stainless steel vacuum chamber maintained at < $10^{-11}$ Torr. Laser-cooled $^{87}$Rb atoms are loaded into a 1-D optical lattice created by a retro-reflected Yb fiber laser beam. The trap beam is focused to a minimum waist of $w_0$ = 12.7 μm, which produces an optical dipole trapping potential of $U_{dipole}$ = 2 mK for 1 W of power. A six-beam magneto-optical trap (MOT) is used to provide the sample of laser-cooled atoms. The laser cooling beams are detuned −16 MHz from the $5S_{1/2}$ $F$ = 2 → $5P_{3/2}$ $F'$ = 3 transition of $^{87}$Rb and have an intensity of 2.4 mW/cm$^2$ per beam. The magnetic field gradient for the MOT is created by two anti-Helmholtz magnetic coils. To load single atoms, a magnetic field gradient of 350 G/cm is utilized. For larger atomic samples, the field gradient is lowered to 18 G/cm in order to expand the trapping volume. Loading the laser cooled atoms into the optical lattice is accomplished by operating the MOT and lattice concurrently. After 10 s of loading, the magnetic field is turned off, leaving some of the atoms trapped in the lattice.

The trapped atoms are detected by imaging atomic fluorescence onto an electron multiplying CCD camera. The atoms are excited using the laser cooling beams. For trap depths of >1 mK, and with careful alignment and balance of the laser beams, the atoms can be simultaneously cooled and observed non-destructively [16]. The atomic fluorescence is collected with a high numerical aperture (NA = 0.40) microscope objective. With this imaging system, it is possible to detect individual atoms, both in the MOT and in the optical lattice, with exposure times as short as 100 ms. Fig. 1(a) shows an image of a sparsely loaded optical lattice. A typical evolution of the trap population versus time is shown in Fig. 1(b). For this data, the atoms were continuously cooled and monitored. Note that one of the atoms remains trapped for more than 600 s. An exponential fit to this limited data set indicates an 1/$e$ lifetime of 270 s, which is consistent with a vacuum-limited lifetime at our measured pressure of $\sim 10^{-11}$ Torr.

Surprisingly, we find that the lifetime of the atoms in the lattice is dramatically lower when the cooling light is not applied. The decay of the trapped atoms in this case is shown in Fig. 2. Each data point is the average of 5 runs and the lattice is reloaded for each run because atom counting is destructive in this case. The total elapsed time for the data in this figure is 10 hrs. For this data, the initial number of atoms in the trap was ~500-1000.

In addition to the shorter overall decay time, the uncooled atoms exhibit a non-exponential decay. Following loading, there is an initial period of ~20 s during which the atom loss is minimal, as shown in the inset to Fig. 2. Subsequently, the population decays exponentially with a 62 s time constant. This behavior is consistent with a heating source which continuously increases the total energy of the atoms. The initial delay in the atom loss is due to the fact that the thermal energy of the atoms immediately after loading is considerably below the well depth. Thus, there is a time delay before atoms gain enough energy to leave the trap.

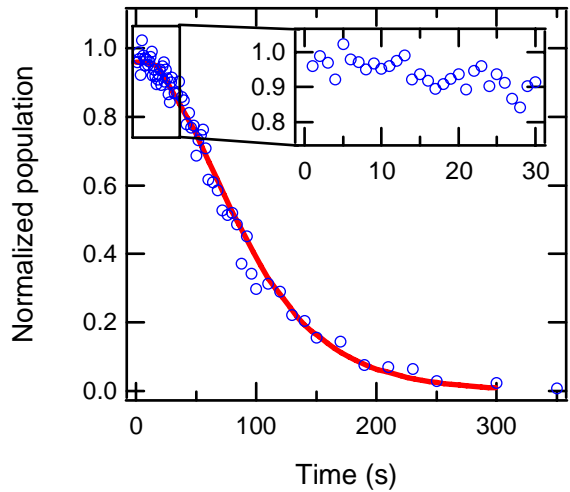

FIG. 2: Atom population in the optical lattice vs. time with no cooling. The asymptotic decay of the trap is exponential with a 62 s lifetime. Inset shows the trap population for the first 30 s. The solid line is a simulation including heating as discussed in the text.

In order to quantitatively compare our results to possible heating sources, we determine the heating rate in the trap derived from measurements of the intensity noise power spectra [10] of the trap beam and the noise power spectra for fluctuations of the trap equilibrium position. Fluctuations in the laser intensity cause parametric heating of the atoms due to a modulation of the trapping potential. This leads to an exponential energy growth of the atoms with a time constant Γ given by [10],

$$\Gamma = \frac{1}{T_I} = \pi^2 \nu_{tr}^2 S_I(2\nu_{tr}), \quad (1)$$

where $T_I$ is the energy $e$-folding time, $\nu_{tr}$ is the trap frequency, and $S_I(2\nu_{tr})$ is the relative intensity noise power spectrum. The other dominant heating mechanism results from fluctuations in the trap position (*e.g.* due to laser beam pointing or phase instabilities). In this case, the energy grows linearly with a heating rate $\dot{Q}$ given by,

$$\dot{Q} = 4\pi^4 \nu_{tr}^4 m S_x(\nu_{tr}) \quad (2)$$

where $S_x(\nu_{tr})$ is the trap position noise power spectrum.

We measure the intensity noise and the position noise in the radial direction of the trapping beam using a balanced detection method [17]. The position noise in the axial direction is measured using an interferometric technique. The corresponding heating rate $\dot{Q}$ and heating time constant $\Gamma$ calculated from Eqs. (1) and (2) are shown in Fig. 3.

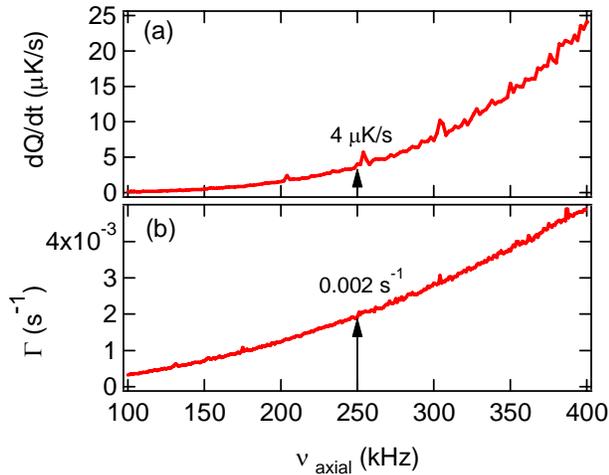

FIG. 3: (a) Heating rate due to fluctuations in the axial trap position. (b) Heating time constant due to intensity fluctuations.

The axial and radial trapping frequencies of the optical lattice are measured using parametric excitation [18], resulting in values of $\nu_{axial} = 250$ kHz and $\nu_{radial} = 2.8$ kHz respectively. The corresponding $\dot{Q}$ and $\Gamma$ for the axial direction are 4 μK/s and 0.002 s$^{-1}$, respectively. The heating rates in the radial directions are negligible compared to the heating in the axial direction because these heating processes scale as $\nu_{tr}^2$ and $\nu_{tr}^4$ and are therefore not shown in the figure.

The time evolution of the trap population can be modeled with a Fokker-Planck equation for the energy distribution $n(E,t)$ given by [19],

$$\frac{\partial n}{\partial t} = \left(\frac{\Gamma}{4}E^2 + \dot{Q}E\right)\frac{\partial^2 n}{\partial E^2} - \dot{Q}\frac{\partial n}{\partial E} - \frac{\Gamma}{2}n. \quad (3)$$

Numerical solutions of Eq. (3) are obtained assuming an initial Maxwell-Boltzmann distribution with a temperature of the trapped atoms of 100 μK. The simulation results, shown as a solid curve in Fig. 2 for the following parameters, $\dot{Q} = 4.5$ μK/s, $\Gamma = 0.002$ s$^{-1}$, closely reproduce the observed trap population, and in particular, show the 20 s delay before the onset of appreciable trap loss.

The remarkable feature of our observation, which is supported by the simulation, is that it takes a finite amount of time for the atoms to heat up sufficiently to be ejected from the trap. It follows that it should be possible to extend the lifetime of the trapped atoms by occasionally re-cooling them to the bottom of the trap. In Fig. 4, we demonstrate the proof of principle of this idea by applying a short 5 ms pulse of laser cooling light to the atoms at $t = 15$ s. The cooling light is provided by the laser beams used to form the MOT. As is evident from the data, the loss of atoms is halted for a time comparable to the delay in atom loss following initial loading. Note for this data, the overall lifetime is shorter than in Fig. 2. This is due to the fact that for this data, a different Yb fiber laser with a much higher intensity noise is used as the trapping laser.

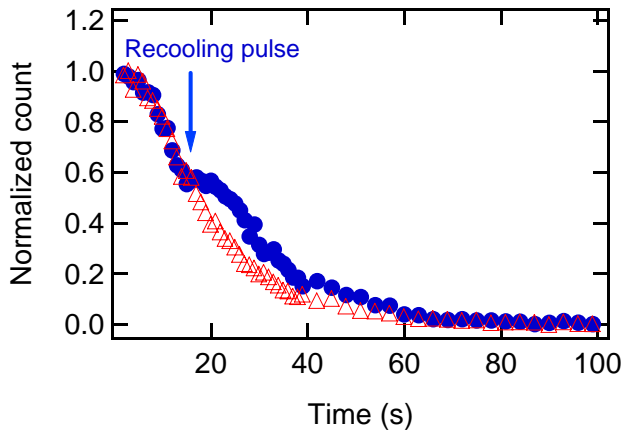

FIG. 4: (color online) (●) Lattice population vs. time showing the effect of a single 5 ms cooling pulse applied at $t = 15$ s. (△) Similar data without the cooling pulse applied.

The results in Fig. 4 suggest that it should be possible to minimize atom loss caused by heating almost entirely by providing cooling pulses at time intervals shorter than the initial heating time. In Fig. 5, we demonstrate that this is indeed possible and realize a dramatic increase in the lifetime by application of a periodic cooling pulse to the atoms. For this data (◊), the atoms are cooled with a 1 s cooling pulse applied every 19 s. The trap population for this cooling method shows a simple exponential decay with a 310 s lifetime. To our knowledge, this is the longest lifetime reported in an optical lattice by a factor of 5 [15].

We have compared the lifetime of the pulsed cooling method to the case where the atoms are continuously exposed to the cooling light. This is also shown in Fig. 5. The continuously cooled atoms exhibit an initial fast decay rate ($\Gamma_{initial}^{-1} = 45$ s) in the first ~100 s followed by a slower decay at approximately the same rate as the pulsed-cooling case and the rate inferred from the distinguishable trapped atom case shown in Fig. 1. We attribute this fast initial decay to light assisted collisions between two atoms trapped in the same anti-node. For this data, the initial loading was ~700 atoms distributed over ~500 sites, and hence one expects >40% of the sites to have at least 2 atoms initially assuming Poissonian loading statistics. Two atoms per site corresponds to an effective density of $n = 1.1 \times 10^{10}$ cm$^{-3}$ at the doppler temperature. The loss rate due to light assisted collisions at this density for the intensity and detuning of our cooling beams is $\Gamma_{light} = 0.02$ s$^{-1}$ [20, 21], which is consistent with the fast initial loss rate for the continuous cooling data in Fig. 5. Once the multiply occupied sites are vacated, the decay rate reduces to the pulsed cooling case, which is likely limited by the background vacuum. Note that the pulsed cooling data also has a similar number of sites with multiple atoms. In this case, however, the light is on for only 1/20 of the time, thus the effective $\Gamma_{light}$ is correspondingly reduced.

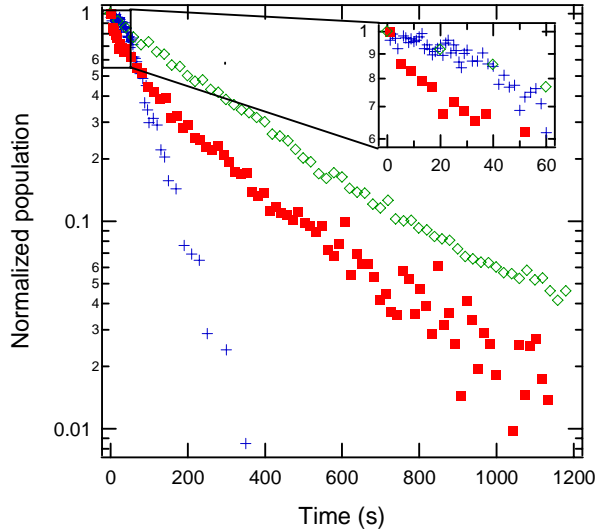

FIG. 5: (color online) Number of atoms in the trap vs. time, normalized to the initial loading for three difference cases: (◊) pulse-cooled atoms, (■) continuously-cooled atoms, and (+) uncooled atoms. The inset shows the first 60 s of data.

The trap lifetime in the absence of cooling is limited by fluctuations in the axial trap position due to vibrations in the mount for the mirror used to form the lattice standing wave. Although the measured fluctuations of the lattice standing wave are very small, ($\Delta x_{rms} \sim 10^{-4} \lambda$ in the frequency range from 10 kHz to 2 MHz), they provide the dominant heating source. In the current set-up, it is necessary to have an adjustable mirror mount to achieve the required alignment of the lattice. It should be straightforward to reduce these vibrations by moving to a fixed mount in the future. The heating due to the laser intensity noise will limit the lifetime to ~1000 s, which is much longer than the background-limited lifetime.

In conclusion, we have examined lifetime limitations due to heating in a long-lived 1-D optical lattice. We have been able to identify the dominant heating mechanism in the trap by comparing our results with measured fluctuations of the trap parameters. Furthermore, we have demonstrated that it is possible to greatly extend the trap lifetime in our optical lattice through judicious application of laser cooling to counter-act the heating. With this, we were able to extend the lifetime beyond 300 s. The results of this study can be employed in designing optical lattices with very long lifetimes, which is an important requirement for many current research applications.

We would like to acknowledge valuable discussions with Paul Griffin. This work is supported by the National Science Foundation, PHYS-0605049 and PHYS-0703030.